\documentclass[twoside,12pt]{article}
\usepackage{epsfig}

\def\Journal#1#2#3#4{{#1} {#2} (#4) #3 }

\def\PLB{{\em Phys. Lett.} B}

\def\PRL{\em Phys. Rev. Lett.}

\def\PRD{{\em Phys. Rev.} D}

\def\INTA{{\em Int. J. Mod. Phys.} A} 
\def\NPPS{\em Nucl. Phys. Proc. Suppl.} 
\def\AIP{\em  AIP Conf. Proc.}

\newcommand{\be}{\begin{equation}}
\newcommand{\ee}{\end{equation}}
\newcommand{\bea}{\begin{eqnarray}}
\newcommand{\eea}{\end{eqnarray}}
\newcommand{\eq}{\begin{eqnarray}}
\newcommand{\en}{\end{eqnarray}}

\newcommand{\la}{\langle}
\newcommand{\ra}{\rangle}
\newcommand{\bfb}{{\bf b}_{\perp}}

\begin{document}

\title{ \vspace{1cm} Hadron properties in AdS/QCD}
\author{\hspace*{-1cm} Thomas Gutsche$^1$, Valery E. Lyubovitskij$^1$
\footnote{ 
On leave of absence from Department of Physics, Tomsk State University, 
634050 Tomsk, Russia}, 
Ivan Schmidt$^2$, Alfredo Vega$^2$ 
\\
$^1$Institut f\"ur Theoretische Physik,
    Universit\"at T\"ubingen,\\
    Kepler Center for Astro and Particle Physics, \\
    Auf der Morgenstelle 14, D-72076 T\"ubingen, Germany\\
$^2$Departamento de F\'\i sica y Centro Cient\'\i fico
    Tecnol\'ogico de Valpara\'\i so (CCTVal),\\
    Universidad T\'ecnica Federico Santa Mar\'\i a,\\
    Casilla 110-V, Valpara\'\i so, Chile
}

\maketitle
\begin{abstract} 
We discuss a holographic soft-wall model developed for the
description of mesons and baryons with adjustable quantum numbers
$n, J, L, S$. This approach is based on an action which describes
hadrons with broken conformal invariance and which incorporates
confinement through the presence of a background dilaton field.
Results obtained for heavy-light meson
masses and decay constants are consistent with predictions of HQET.
In the baryon sector applications to the baryon masses, 
nucleon electromagnetic form factors and generalized parton 
distributions are discussed. 
\end{abstract} 
\section{Introduction}
\label{intro}

Based on the correspondence of string theory in anti-de Sitter (AdS)
space and conformal field theory (CFT) in physical
space-time~\cite{Maldacena:1997re}, a class of AdS/QCD approaches
was recently successfully developed for describing the phenomenology
of hadronic properties. In order to break conformal invariance and
incorporate confinement in the infrared (IR) region two alternative
AdS/QCD backgrounds have been suggested in the literature: the
``hard-wall'' approach~\cite{Hard_wall1}, based on
the introduction of an IR brane cutoff in the fifth dimension, and
the ``soft-wall'' approach~\cite{Soft_wall1}, based on
using a soft cutoff. This last procedure can be introduced in the
following ways: i) as a background field (dilaton) in the overall
exponential of the action, ii) in the warping factor of the AdS
metric, iii) in the effective potential of the action. These methods
are in principle equivalent to each other due to a redefinition of
the bulk field involving the dilaton field or by a
redefinition of the effective potential.
In the literature there exist detailed discussions of the sign
of the dilaton profile in the dilaton exponential
$\exp(\pm\varphi)$~\cite{Soft_wall1,Soft_wall2a,Soft_wall2b,%
SWminus,SWplus} for the soft-wall model (for a discussion of the
sign of the dilaton in the warping factor of the AdS metric see
Refs.~\cite{Soft_wall3a}). The negative sign was suggested in
Ref.~\cite{Soft_wall1} and recently discussed in Ref.~\cite{SWplus}.
It leads to a Regge-like behavior of the meson spectrum, including a
straightforward extension to fields of higher spin~$J$. Also, in
Ref.~\cite{SWplus} it was shown that this choice of the dilaton sign
guarantees the absence of a spurious massless scalar mode in the
vector channel of the soft-wall model. We stress that alternative
versions of this model with a positive sign are also possible. One
should just redefine the bulk field $S(x,z)$ as 
$S(x,z) = e^{\varphi(z)} \, \tilde S(x,z)$, 
where the transformed field corresponds to the dilaton with an opposite
profile. It is clear that the underlying action changes, and extra
potential terms are generated depending on the dilaton field
(see detailed discussion in~\cite{Gutsche:2011vb}).
Here we present a summary of recent results: 
meson mass spectrum and decay constants of light and heavy mesons, 
baryon masses, nucleon electromagnetic form factors and generalized parton  
distributions~\cite{Gutsche:2011vb}-\cite{Soft_wall9}.   
Our starting point are the 
effective $(d+1)$ dimensional actions formulated in AdS space 
in terms of boson or fermion bulk fields, which serve as holographic 
images of mesons and baryons.

\section{Approach} 

Our starting point are the 
effective $(d+1)$ dimensional actions formulated in AdS space 
in terms of boson or fermion bulk fields, which serve as holographic 
images of mesons and baryons. For illustration we consider the 
simplest actions --- for scalar fields 
($J=0$)~\cite{Soft_wall8}-\cite{Soft_wall5} 
\eq 
S_0 = \frac{1}{2} \int d^dx dz \sqrt{g} e^{-\varphi(z)}
\biggl[ g^{MN} \partial_M S(x,z) \partial_N S(x,z)
- \Big(\mu_S^2 + \Delta V_0(z)\Big) \, S^2(x,z) \biggr] \,. 
\en
and $J=1/2$ fermions~\cite{Abidin:2009hr,Soft_wall9}: 
\eq
S_{1/2} &=&  \int d^dx dz \, \sqrt{g} \, e^{-\varphi(z)} \,
\biggl[ \frac{i}{2} \bar\Psi(x,z) \epsilon_a^M \Gamma^a
{\cal D}_M \Psi(x,z) - \frac{i}{2} ({\cal D}_M\Psi(x,z))^\dagger \Gamma^0
\epsilon_a^M \Gamma^a \Psi(x,z) \nonumber\\
&-& \bar\Psi(x,z) \Big(\mu_\Psi + \varphi(z)/R)\Big)\Psi(x,z)\biggr] \,, 
\en 
where $S$ and $\Psi$ are the scalar and fermion bulk fields, 
${\cal D}_M$ is the covariant derivative acting on the fermion field, 
$\Gamma^a=(\gamma^\mu, - i\gamma^5)$ are the Dirac matrices, 
$\varphi(z) = \kappa^2 z^2$ is the dilaton field, $R$ is the AdS radius, 
$\Delta V_0(z)$ is the dilaton potential. 
$\mu_S$ and $\mu_\Psi$ are the masses of scalar and fermion bulk fields 
defined as $\mu_S^2 R^2 = \Delta_M (\Delta_M - d)$ and 
$\mu_\Psi R = \Delta_B - d/2$. Here 
$\Delta_M = \tau_M = 2 + L$ and $\Delta_B = \tau_B + 1/2 = 7/2 + L$
are the dimensions of scalar and fermion fields, which due to 
the QCD/gravity correspondence are related to the scaling dimensions 
(twists $\tau_M, \tau_B$) of the corresponding interpolating operators, 
where $L = {\rm max} \, | L_z |$~\cite{Soft_wall2a} is the maximal 
value of the $z$-component of the quark orbital angular momentum 
in the LF wavefunction or the minimum of the orbital angular momentum 
of the corresponding hadron. 
These actions give information about the propagation of bulk fields 
inside AdS space (bulk-to-bulk propagators), from inside 
to the boundary of the AdS space (bulk-to-boundary propagators) 
and bound state solutions - profiles of the Kaluza-Klein (KK) modes in 
extra-dimension, which correspond to the hadronic wave functions 
in impact space. We suppose a free propagation
of the bulk field along the $d$ Poincar\'e coordinates with four-momentum
$p$, and a constrained propagation along the $(d+1)$-th coordinate $z$
(due to confinement imposed by the dilaton field). 
In particular, it was shown~\cite{Soft_wall2a}   
that the extra-dimensional coordinate $z$ corresponds 
to the light-front impact variable. It was also shown~\cite{SWplus}  
that in case of the scattering problem the sign of the dilaton profile is 
important to fulfill certain model-independent constraints.  
But we recently showed~\cite{Gutsche:2011vb} that in case of the 
bound state problem the sign of the dilaton profile is irrelevant, 
if the action is properly set up. Moreover, in solving the bound-state 
problem, it is more convenient to move the dilaton field from the exponential 
prefactor to the effective potential~\cite{Soft_wall2a,Gutsche:2011vb}.  
Then we use a KK expansion for the bulk fields factorizing 
the dependence on $d$ Poincar\'e coordinates $x$ and the holographic 
variable $z$. E.g. in case of scalar field it is given by    
$S(x,z) = \sum_n \ S_n(x) \ \Phi_{n}(z)$, 
where $n$ is the radial quantum number, $S_n(x)$
is the tower of the KK modes dual to
scalar mesons and $\Phi_n$ are their extra-dimensional profiles
(wave-functions) satisfying the Schr\"odinger-type equation with 
the potential depending on the dilaton field. 
Then using the obtained wave functions $\Phi_n$ we calculate matrix elements 
describing hadronic processes. 

\section{Results} 

We present the results of our calculations for mesonic decay 
constants (Table 1), the meson spectrum (Tables 2 and 3)~\cite{Soft_wall8} 
and the baryon spectrum (Tables 4 and 5). 
A detailed analysis of nucleon helicity-independent generalized parton 
distributions (GPDs) $H_{v}^{q}$ and $E_{v}^{q}$~\cite{Soft_wall9} is 
discussed.  
Note, by construction we reproduce the power scaling of nucleon 
electromagnetic (EM) form factors at large 
$Q^2$~\cite{Abidin:2009hr,Soft_wall9}: 
\eq 
F_{1}^p(Q^2) &=& C_{1}(Q^2) + \eta_{p} C_{2}(Q^2)\,,\nonumber\\
F_{2}^p(Q^2) &=& \eta_{p} C_{3}(Q^2)\,,\nonumber\\
& &\\
F_{1}^n(Q^2) &=& \eta_{n} C_{2}(Q^2)\,,\nonumber\\
F_{2}^n(Q^2) &=& \eta_{n} C_{3}(Q^2), \nonumber 
\en 
where $Q^{2} = - t$. The $C_i$ functions, defining the Dirac and Pauli  
factors, are given by:   
\eq
C_1(Q^2) &=& \frac{a+6}{(a+1)(a+2)(a+3)} \,, \nonumber\\
C_2(Q^2) &=& \frac{2a (2a-1)}{(a+1)(a+2)(a+3)(a+4)} \,, \\
C_3(Q^2) &=& \frac{12 m_N \sqrt{2}}{\kappa} \,  
\frac{1}{(a+1)(a+2)(a+3)} \,, \nonumber 
\en 
where $a = Q^2/(4\kappa^2)$. Note that we obtain the correct 
scaling behavior of the nucleon form factors at large $Q^2$, 
$F_1^{p,n} \sim 1/Q^4$ and $F_2^{p,n} \sim 1/Q^6$. 

Our predictions for 
the EM radii compare well with data: 
\eq 
& &\la r^2_E \ra^p =  
0.91 \ {\rm fm}^2 \ {\rm (our)}\,,
\ 
0.77\ {\rm fm}^2 \ {\rm (data)}\, ; \ 
\la r^2_E \ra^n =    
- 0.12 \ {\rm fm}^2 \ {\rm (our)}\,, \ 
- 0.12 \ {\rm fm}^2 \ {\rm (data)}\,, \nonumber\\
& &\la r^2_M \ra^p = 0.85 \ {\rm fm}^2 \ {\rm (our)}\,,
\ 
0.73 \ {\rm fm}^2 \ {\rm (data)}\, ; \  
\la r^2_M \ra^n =  0.88 \ {\rm fm}^2 \ {\rm (our)}\,,
\
0.76 \ {\rm fm}^2 \ {\rm (data)}\,. \nonumber
\en 
In the following we discuss in detail our results for GPDs, which 
are related to the electromagnetic form factors via 
sum rules~\cite{Ji:1996nm,Radyushkin:1997ki}. 
In particular, in momentum space $H_{v}^{q}$ and $E_{v}^{q}$ are given by 
\eq 
H_{v}^{q}(x,Q^2) &=& q(x) \, x^a \,, 
\label{HqAdS}\\ 
E_{v}^{q}(x,Q^2) &=& e^q(x) \, x^a \,. 
\label{EqAdS}  
\en 
Here $q(x)$ and $e^q(x)$ are distribution functions given by: 
\eq 
q(x)   = \alpha^q \gamma_{1}(x) + \beta^q \gamma_{2}(x)\,, \quad 
e^q(x) = \beta^q \gamma_{3}(x)\,, 
\en 
where the flavor couplings $\alpha^q, \beta^q$ 
and functions $\gamma_i(x)$ are written as
\eq 
\alpha^u = 2\,, \ \alpha^d = 1\,, \ 
\beta^u = 2 \eta_{p} + \eta_{n} \,, \ 
\beta^d = \eta_{p} + 2 \eta_{n} \,  
\en 
and 
\eq
\gamma_{1}(x) &=& 
\frac{1}{2} (5 - 8x + 3x^{2})\,, \nonumber\\
\gamma_{2}(x) &=& 1 - 10x + 21x^{2} - 12x^{3} \,, 
\label{gamma} \\
\gamma_{3}(x) &=& 
\frac{6 m_N \sqrt{2}}{\kappa} (1 - x)^{2} \,. \nonumber
\en 
The parameters are $\kappa = 350$ MeV, $\eta_{p} = 0.224$,
$\eta_{n} = -0.239$, which were
fixed in order to reproduce the mass $m_N = 2\kappa \sqrt{2}$ and 
the anomalous magnetic moments of the nucleon 
$k_p = \mu_p - 1 = 1.791$ and $k_n = \mu_n = - 1.913$. The plots of 
the GPDs in momentum space are presented in Fig.1. 

Another interesting aspect to consider are the nucleon GPDs 
in impact space. As shown in~\cite{Burkardt:2000za}, 
the GPDs in momentum space are related to the impact parameter 
dependent parton distributions by a Fourier transform. 
GPDs in impact space give access to 
the distribution of partons in the transverse plane, which is 
quite important for understanding the nucleon structure.
Soft-wall AdS/QCD gives the following predictions for 
the impact space properties of nucleons: 
\eq 
q(x,\bfb) &=& q(x) \frac{\kappa^2}{\pi\log(1/x)} 
e^{- \frac{\bfb^2\kappa^2}{\log(1/x)}} \,, \nonumber\\
e^q(x,\bfb) &=& e^q(x) \frac{\kappa^2}{\pi\log(1/x)} 
e^{- \frac{\bfb^2\kappa^2}{\log(1/x)}} \,, \nonumber\\ 
\rho_E^N(\bfb) &=& \frac{\kappa^2}{\pi}
\sum\limits_{q} e_q^N \int\limits_0^1 \frac{dx}{\log(1/x)}
q(x) e^{- \frac{\bfb^2\kappa^2}{\log(1/x)}} \,, \nonumber\\
\rho_M^N(\bfb) &=& \frac{\kappa^2}{\pi}
\sum\limits_{q} e_q^N \int\limits_0^1 \frac{dx}{\log(1/x)}
e^q(x) e^{- \frac{\bfb^2\kappa^2}{\log(1/x)}} \,, \nonumber\\ 
\la R_\perp^2(x) \ra_q &=& \frac{\log(1/x)}{\kappa^2} \,, \nonumber\\
\la R_\perp^2\ra_q &=& \frac{1}{\kappa^2}
\biggl( \frac{5}{3} + \frac{\beta^q}{12 \alpha^q} \biggr) \,. 
\en  
Fig.3 shows some examples of $q(x,\bfb)$ and in Fig.4 
we plot $\rho_E^N(\bfb)$ and $\rho_M^N(\bfb)$ for proton and neutron. 
For the transverse rms radius of $u-$ and $d-$quark GPDs we get 
similar values: 
\eq 
\la R_\perp^2\ra_u = 0.527 \ {\rm fm}^2\,, \quad 
\la R_\perp^2\ra_d = 0.524 \ {\rm fm}^2\,.  
\en 
One should stress that the obtained nucleon GPDs both in momentum and 
impact spaces correspond to the so-called ``Gaussian ansatz'' and 
are consistent with general predictions for their asymptotic  
behavior for $x \to 0$ or $x \to 1$ and $Q^2 \to 0$ or 
$Q^2 \to \infty$. The plots of GPDs in impact space are given in~Fig.2.  

\begin{center} 

Table 1. Decay constants $f_P$ (MeV) of pseudoscalar mesons 

\vspace*{.25cm}

\def\arraystretch{1.1}
\begin{tabular}{|l|c|c|}
\hline
Meson &Data & Our \\ \hline
$\pi^-$& $130.4\pm 0.03 \pm 0.2$ & 131 \\
$K^-$ & $156.1\pm 0.2 \pm 0.8$ & 155 \\
$D^+$ & $206.7 \pm 8.9$ & 167 \\
$D_s^+$ & $257.5\pm6.1$ & 170 \\
$B^-$&$193\pm11$ & 139 \\
$B_s^0$&$253 \pm 8 \pm 7$ & 144 \\
\hline
\end{tabular}
\end{center}

\newpage 

\begin{center}

Table 2. Masses of light mesons in MeV 

\vspace*{.2cm}

\def\arraystretch{1.1}
\begin{tabular}{|l|c|c|c|l|l|l|l|}
\hline
Meson&$n$&$L$&$S$&\multicolumn{4}{c|}{Mass} \\
\hline
$\pi$&0&0,1,2,3&0&$ 140$&$ 1355$ & $1777$&$2099$ \\ 
$\pi$&0,1,2,3&0&0&$140$&$1355$&$1777$&$2099$ \\ 
$K$& 0&0,1,2,3&$\;0\;$&$496$&$1505$&$1901$ & $2207$ \\ 
$f_0[\bar n n]$&0,1,2,3&1&1&$1114$&$1600$&$1952$&$2244$ \\ 
$f_0[\bar s  s]$&0,1,2,3&1&1&$1304$&$1762$&$2093$&$2372$ \\ 
$a_0(980)$&0,1,2,3&1&1&$1114$&$1600$&$1952$&$2372$ \\ 
$\rho(770)$&0,1,2,3&0&1&$804$&$1565$&$1942$&$2240$ \\ 
$\phi(1020)$ &0,1,2,3&0&1&$1019$&$1818$&$2170$&$2447$ \\ 
$a_1(1260)$&0,1,2,3&1&1&$1358$&$1779$&$2101$&$2375$ \\ 
\hline
\end{tabular}
\end{center}

\vspace*{.1cm} 

\begin{center}

Table 3. Masses of heavy-light mesons in MeV 

\vspace*{.2cm}

\def\arraystretch{1.1}
\begin{tabular}{|l|c|c|c|c|c|c|c|c|}
\hline
Meson&$J^{\rm P}$&$n$&$L$&$S$&\multicolumn{4}{c|}{Mass} \\
\hline
$D(1870)$&$0^{-}$&0&0,1,2,3         &0& 1857 & 2435 & 2696 & 2905 \\ 
$D^{\ast}(2010)$&$1^{-}$&0&0,1,2,3  &1& 2015 & 2547 & 2797 & 3000 \\ 
$D_s(1969)$&$0^{-}$&0&0,1,2,3       &0& 1963 & 2621 & 2883 & 3085 \\ 
$D^{\ast}_s(2107)$&$1^{-}$&0&0,1,2,3&1& 2113 & 2725 & 2977 & 3173 \\ 
$B(5279)$&$0^{-}$&0&0,1,2,3         &0& 5279 & 5791 & 5964 & 6089 \\ 
$B^{\ast}(5325)$&$1^{-}$&0&0,1,2,3  &1& 5336 & 5843 & 6015 & 6139 \\ 
$B_s(5366)$&$0^{-}$&0&0,1,2,3       &0& 5360 & 5941 & 6124 & 6250 \\ 
$B^{\ast}_s(5413)$&$1^{-}$&0&0,1,2,3&1& 5416 & 5992 & 6173 & 6298 \\ 
\hline
\end{tabular}
\end{center}

\vspace*{.1cm}

\begin{center} 
Table 4. Masses of light baryons in MeV 

\vspace*{.2cm}

\def\arraystretch{1.1}
    \begin{tabular}{|c|c|c|}
      \hline
      Baryon & Our results & Data    \\
\hline
        $N$    &    939  & 939                   \\
  $\Lambda$    &   1114  & 1116                  \\
  $\Sigma $    &   1180  & 1189                  \\
  $\Xi    $    &   1328  & 1322                  \\
  $\Delta $    &   1232  & 1232                  \\
  $\Sigma^\ast$&   1381  & 1385                  \\
  $\Xi^\ast$   &   1533  & 1530                  \\
  $\Omega $    &   1688  & 1672                  \\
      \hline
    \end{tabular}
\end{center}

\vspace*{.1cm}

\begin{center}
Table 5. Masses of $N$ and $\Delta$ families in MeV 

\vspace*{.2cm} 

\def\arraystretch{1.1}
\begin{tabular}{|c|c|c|}
      \hline
      Baryon & Our results & Data \\
\hline
$N_{1/2^+}(939)$  &  939  & 939                   \\
$N_{1/2^+}(1440)$ &  1372 & $1440^{+30}_{-20}$   \\
$N_{1/2^+}(1710)$ &  1698 & $1710 \pm 30$        \\
$N_{1/2^+}(1880)$ &  1970 &                    \\
$N_{1/2^+}(2100)$ &  2209 &                    \\
$\Delta_{3/2^+}(1232)$ &  1232 & $1232$   \\
$\Delta_{3/2^+}(1600)$ &  1587 & $1600^{+100}_{-50}$   \\
$\Delta_{3/2^+}(1920)$ &  1876 & $1920^{+50}_{-20}$   \\
      \hline
    \end{tabular}
\end{center}

\begin{figure} 
\begin{center}
\includegraphics[scale=0.7,angle=0]{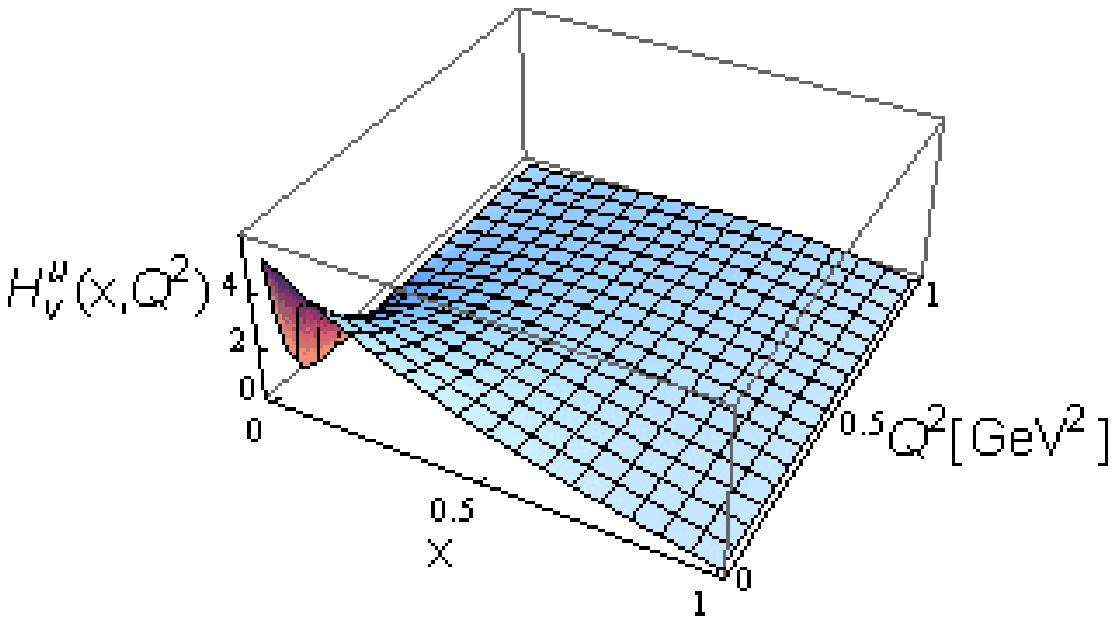} 
\includegraphics[scale=0.7,angle=0]{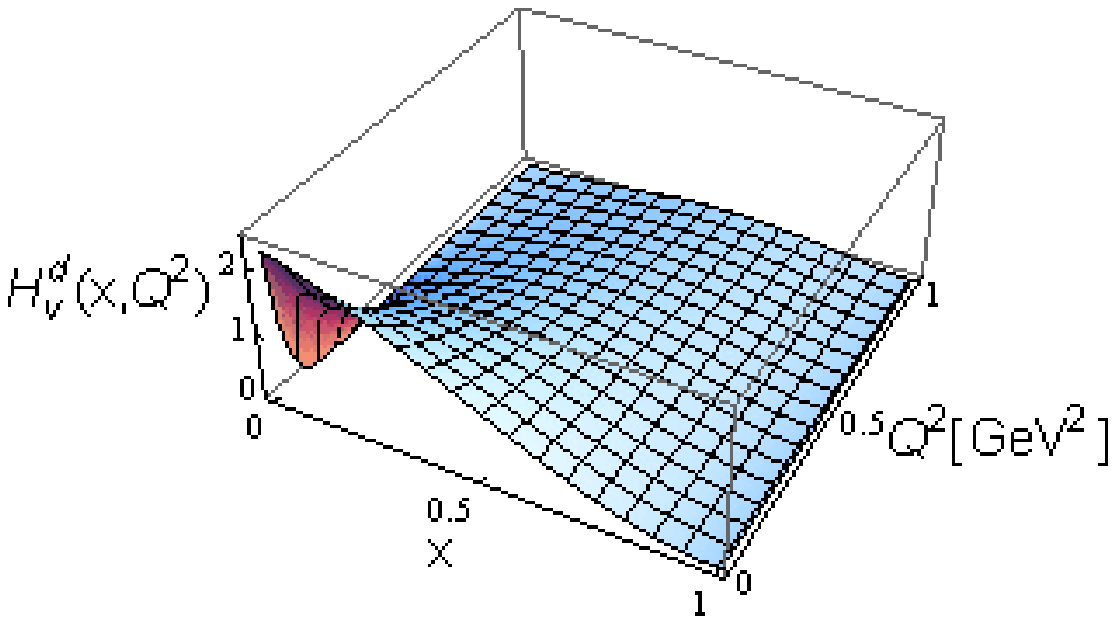} \\
\includegraphics[scale=0.7,angle=0]{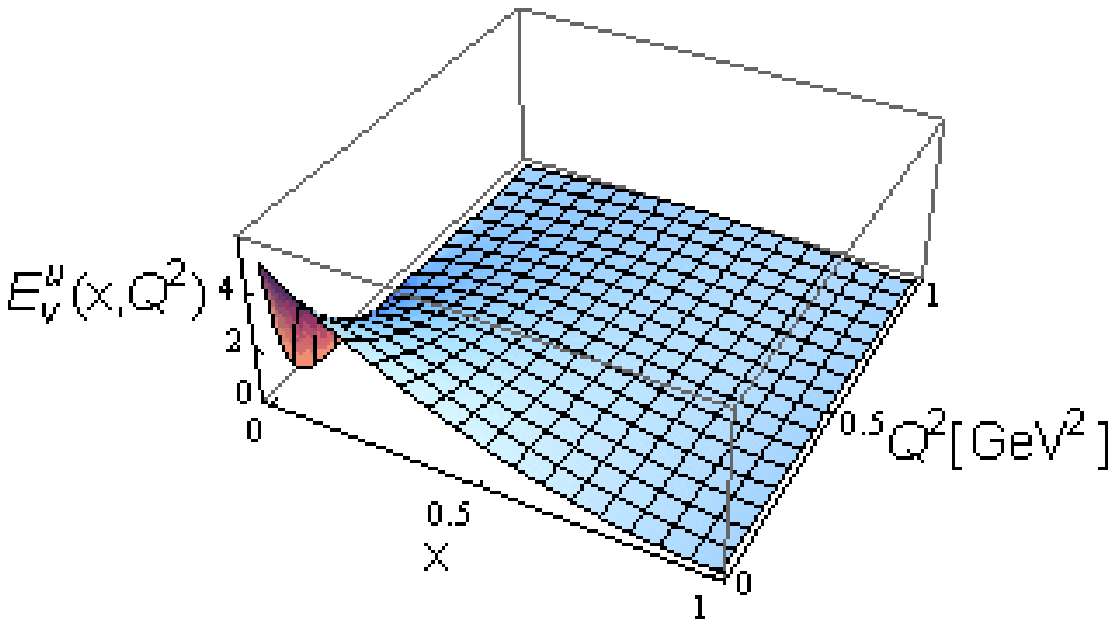} 
\includegraphics[scale=0.7,angle=0]{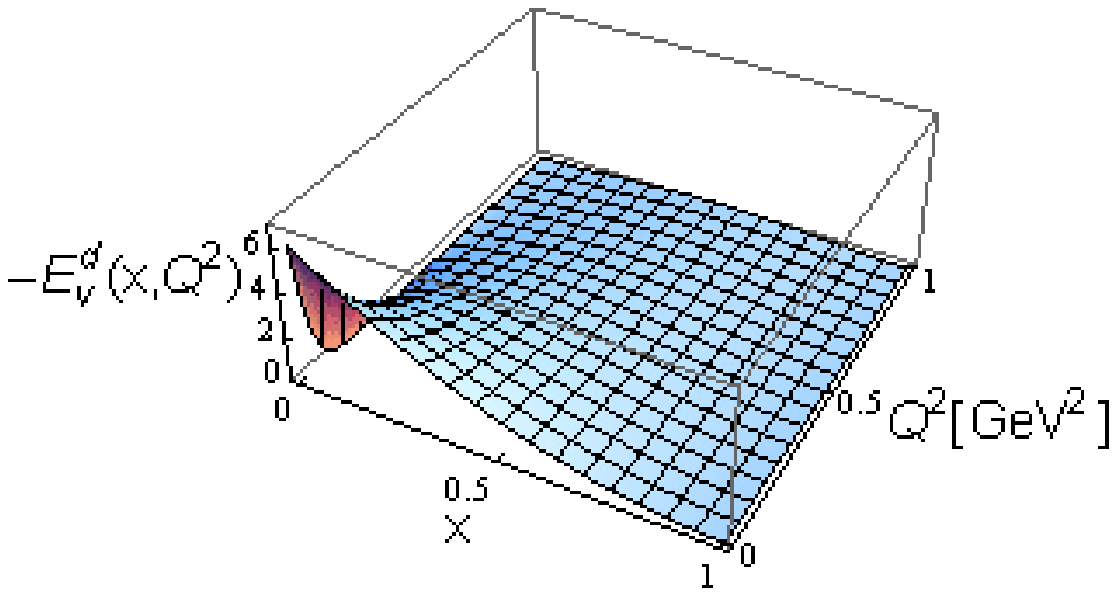} 
\caption{Nucleon helicity-independent GPDs in momentum space} 
\end{center}

\vspace*{.5cm} 

\begin{center}
\begin{tabular}{cc} 
    \includegraphics[scale=0.45,angle=0]{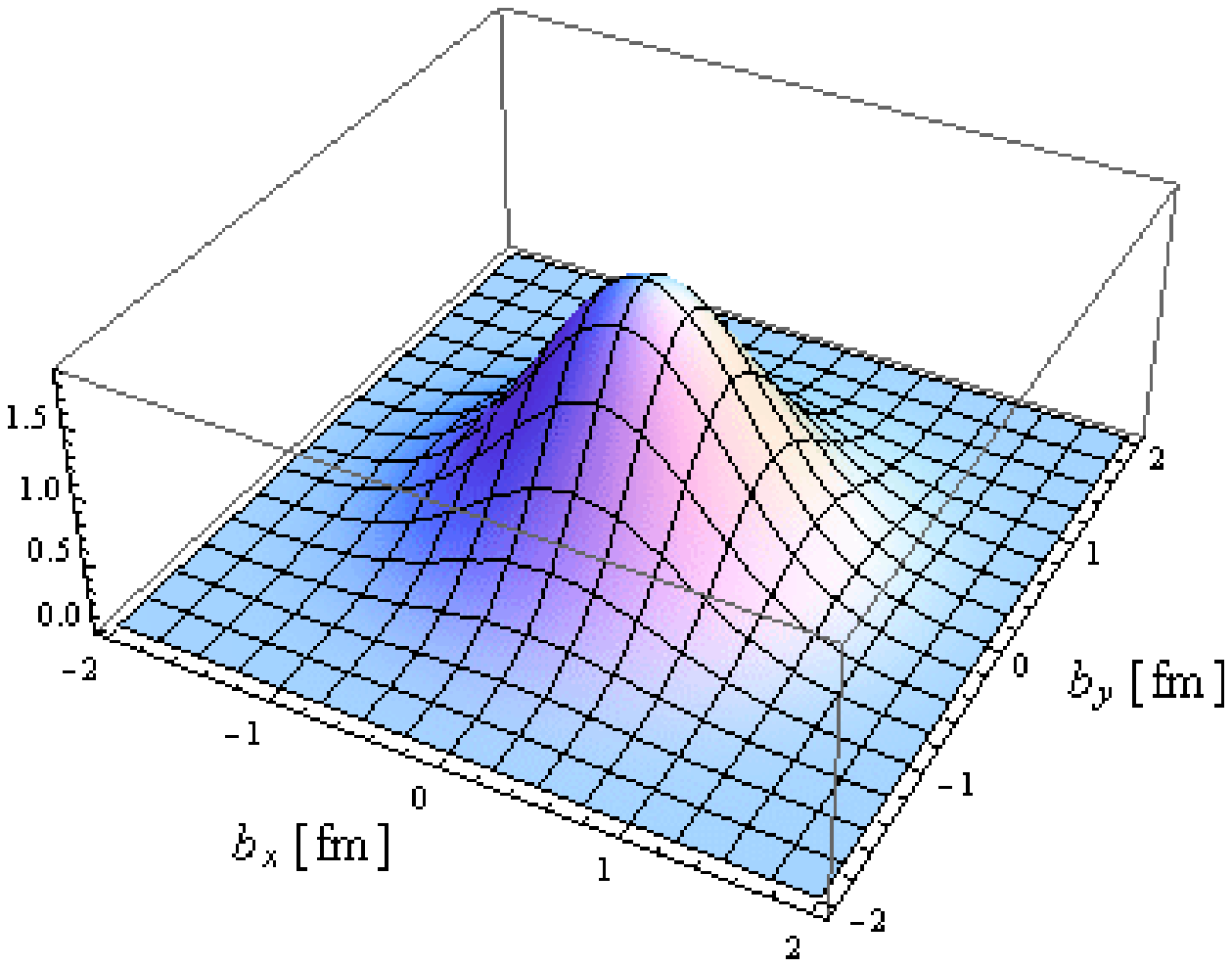} &
    \includegraphics[scale=0.45,angle=0]{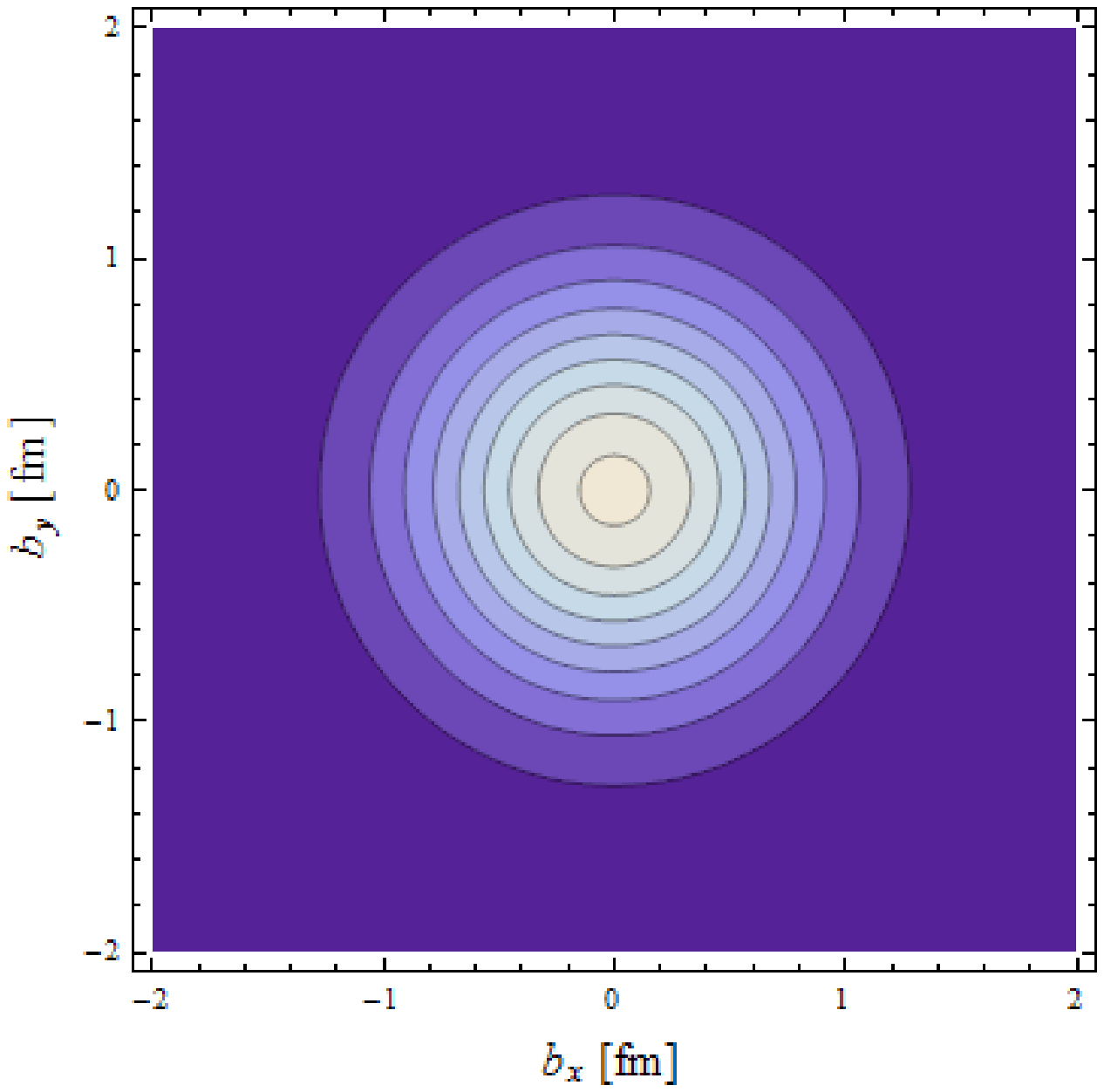} \\
    \includegraphics[scale=0.45,angle=0]{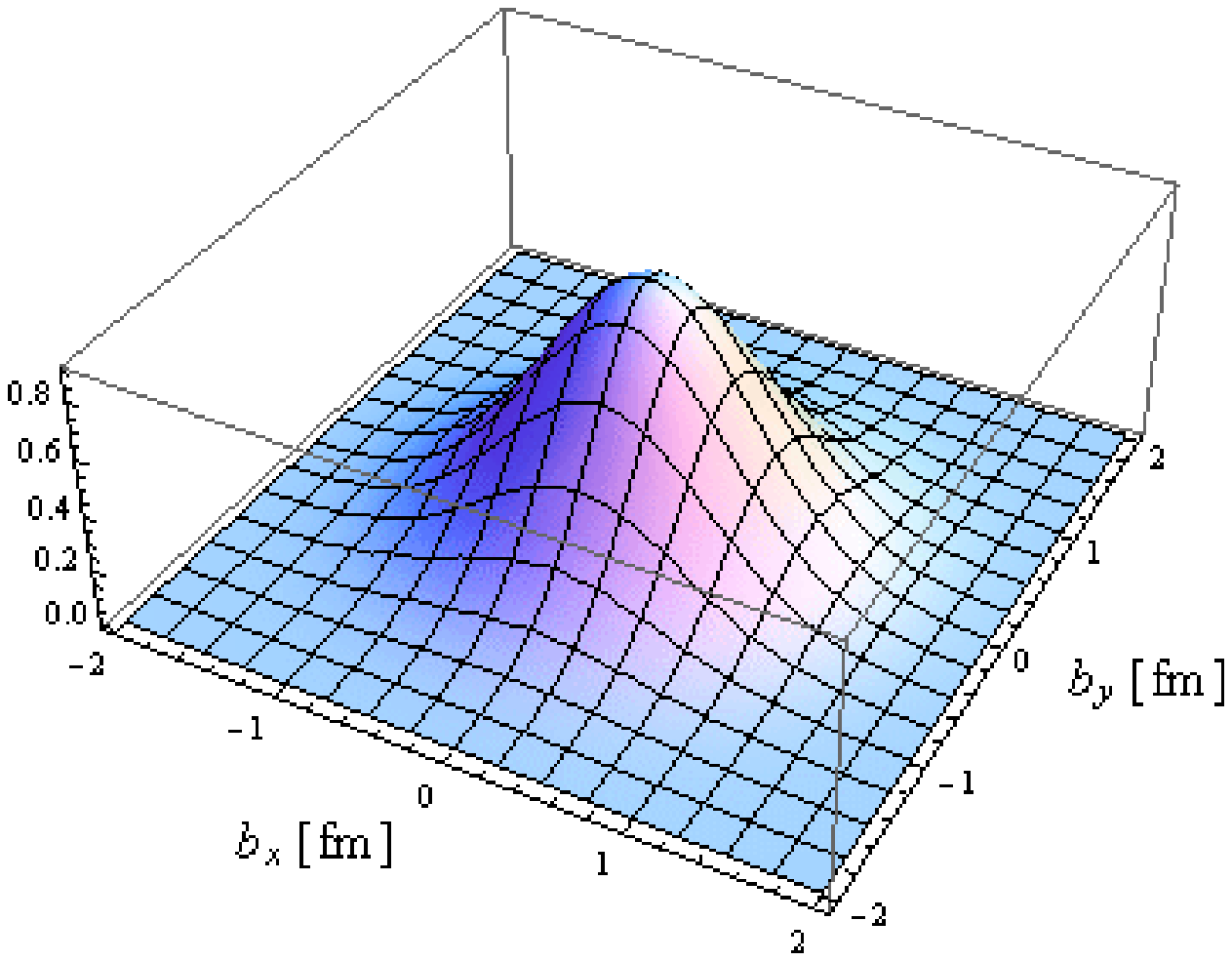} & 
    \includegraphics[scale=0.45,angle=0]{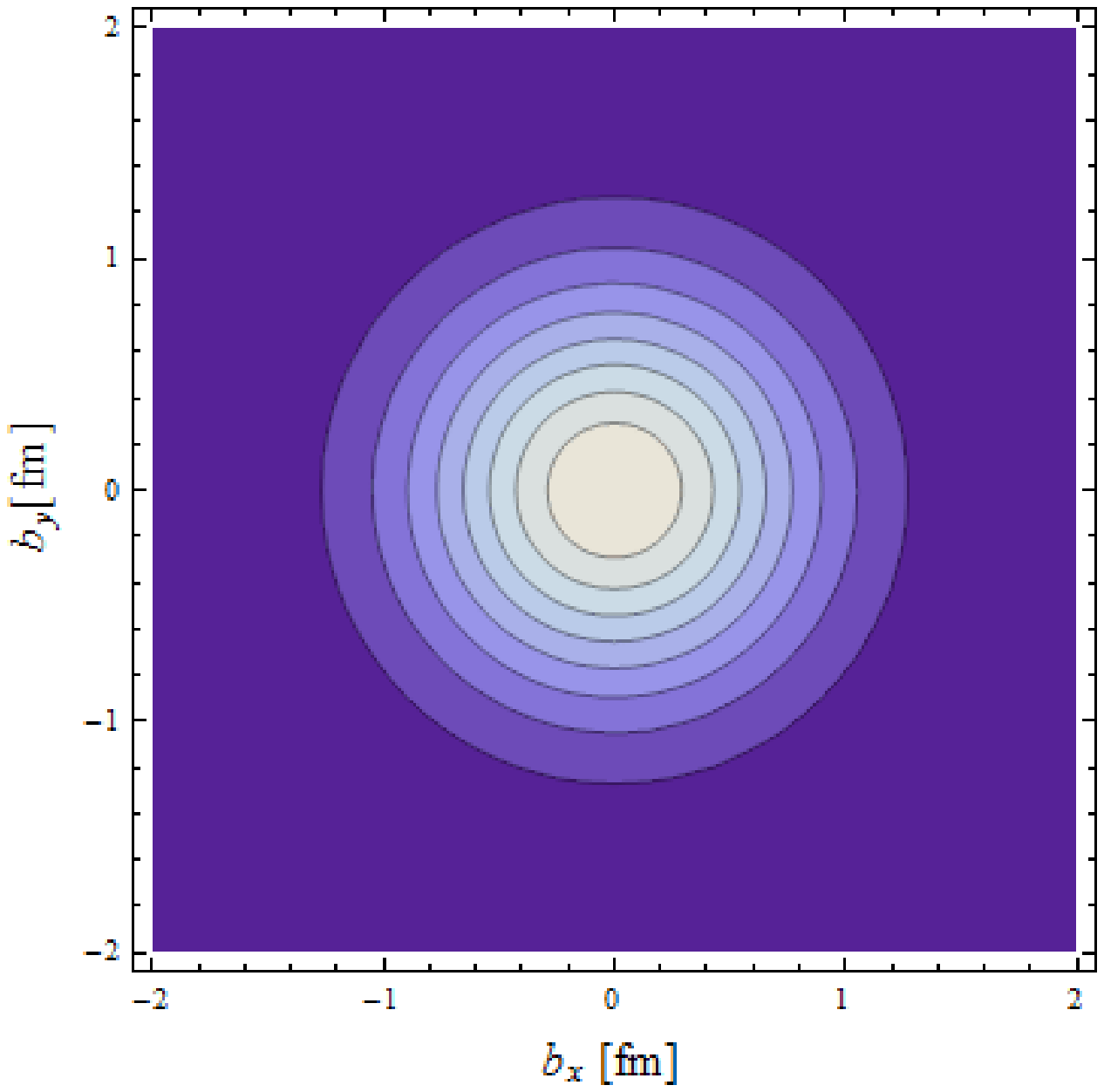}
\end{tabular} 
\caption{Nucleon helicity-independent GPDs in impact space. 
Plots for $q(x,\bfb)$ for $x = 0.1$:
$u(x,\bfb)$ - upper pannels, $d(x,\bfb)$ - lower pannels}
\end{center} 
\end{figure} 

\newpage 

\section{Conclusion}
We present a summary of recent results obtained in a soft-wall model 
based on the gauge/gravity duality:  
meson mass spectrum and decay constants of light and heavy mesons, 
baryon masses, nucleon electromagnetic form factors and generalized 
parton distributions.   

\vspace*{1cm} 

{\large\bf Acknowledgements} 

\vspace*{.5cm} 

The authors thank Stan Brodsky and Guy de T\'eramond for useful discussions
and remarks. This work was supported by Federal Targeted Program ``Scientific
and scientific-pedagogical personnel of innovative Russia''
Contract No. 02.740.11.0238, by FONDECYT (Chile) under Grant No. 1100287. 
A.V. acknowledges the financial support from FONDECYT (Chile)
Grant No. 3100028. 
V.E.L. would like to thank Departamento de F\'\i sica y Centro
Cient\'\i fico Tecnol\'ogico de Valpara\'\i so (CCTVal), Universidad
T\'ecnica Federico Santa Mar\'\i a, Valpara\'\i so, Chile for warm
hospitality.

\end{document}